*Research Article*

# Barter and Hierarchy: A Practical Perspective on Food, Society, and Knowledge in the Inca Empire

*¹Luis-Felipe Arizmendi

## About Article

### About Author

¹ Department of Quantitative Methods, Facultad de Ciencias Económicas y Empresariales, Universidad Pontificia Comillas, Madrid 28015, Spain

### ABSTRACT

The Inca Empire developed a sophisticated system of food production, social organisation, and knowledge transmission without the use of money or writing. This article introduces the concept of a *barter economy structured through hierarchical cooperation*, examining the Inca model from a practice-based perspective. Methodologically, the study draws on ethnohistorical sources, archaeological evidence, and anthropological theory to reconstruct the logic of Inca governance. Key analytical tools include documentary analysis, comparative institutional models, and interpretive examination of empirical innovations such as terracing, *quipus*, and labour systems. Our findings confirm that the Incas achieved food security through ecological complementarity and state-managed redistribution; sustained social cohesion through reciprocity-based labour systems such as the *mita*; and preserved and transmitted knowledge through oral tradition and embodied practices. These results challenge modern assumptions about the necessity of monetary and written systems for complex administration. The Inca case exemplifies a non-market economy organised through stratified reciprocity, where knowledge and cooperation were embedded in daily life and statecraft. The article concludes by reflecting on the relevance of the Inca model for comparative studies in social, economic, and political organisation.

Contact @ Luis-Felipe Arizmendi
lfarizmendi@comillas.edu

## 1. INTRODUCTION

The Inca Empire, known in Quechua as *Tawantinsuyu* ("Realm of the Four Parts"), was the largest pre-Columbian state in the Americas, spanning diverse landscapes from coastal deserts to high Andean plateaus. From the 13th to the 16th century AD, the Incas forged a complex civilisation remarkable for lacking certain features that were standard to other great empires: they had no currency-based market system or written script. Yet, they managed to ensure the sustenance of millions of subjects, construct monumental infrastructure, and govern a vast territory with notable administrative efficiency. This apparent paradox – a highly organised empire without money or writing – motivates the present study. We seek to analyse how the Incas achieved food security, social order, and knowledge transmission through what can be termed a *heuristic-practical approach*: privileging empirical experimentation, communal learning, and pragmatic adaptation over formal theory.

Existing descriptions of the Inca economy often emphasise its redistributive nature and the role of reciprocity and state planning instead of the market. Classic studies have variously characterised the Inca state as a form of "socialist empire" (Baudin, 1961) or as an autarkic command economy rooted in kinship obligations (Murra, 2014; Rostworowski, 2005). These interpretations underline how Inca rulers kept power by controlling labour and provisioning, a system where contributions of goods and work were reciprocated with security and gifts from the state. This article builds on such insights but proposes a refined conceptual framework – a barter economy based on hierarchical cooperation – to capture the specific mechanisms of integration and exchange in Inca society. This framework recognises the vertical organisation of the Inca economy (with the state orchestrating production and distribution) combined with a cooperative ethos of reciprocity at all levels of society.

The central research question is: How did the Inca Empire's systems of food production, social-political organisation, and knowledge transmission function in practice, and what does this reveal about its economic model? We approach this question by examining each of these systems and their interdependencies. The Incas prioritised food sufficiency and labour mobilisation as the foundation of their empire, making these domains ideal for investigating the proposed concept of hierarchical barter cooperation.

## 2. LITERATURE REVIEW

### 2.1. On Inca Food Production and Economic Exchange

Scholars concur that agriculture was the economic bedrock of Inca society, sustained by impressive state coordination across ecological zones. The Incas inherited and expanded an Andean tradition of vertical resource management known as the "vertical archipelago" by Murra, wherein communities established enclaves in multiple altitudinal zones to cultivate diverse crops and mitigate climate risks. This system meant that a single kin group (or *ayllu*) could access highland and lowland products, ensuring dietary variety and resilience. For instance, an *ayllu* might farm potatoes and quinoa in the high Andes while tending corn (maise) or coca leaves in lower valleys, trading internally between ecological niches. Colonial-era observers noted the extraordinary range of crops under Inca rule, as said, maise, potatoes, quinoa, coca, beans, squash, peanuts, cotton, and fruits – made possible by careful environmental adaptation (Antúnez de Mayolo, 2011; Guzmán Barrón, 1955). Livestock, especially llamas and alpacas, were likewise managed as part of this integrated agro-pastoral system, valued for transport, wool, and meat.

Crucially, the Inca state took an active role in agricultural management. Terracing (Castro *et al.*, 2019) and irrigation works (Sieczkowska *et al.*, 2022) were constructed or improved on a massive scale to expand arable land and conserve soil and water. Recent studies have highlighted the engineering prowess behind Inca agricultural infrastructure – for example, the optimal design of terraced field retaining walls and sophisticated canal systems for mountain water management. Erickson (1993) documents how pre-Hispanic communities organised labour to build and maintain raised fields and terraces, underscoring that such projects were social as much as technical endeavours. The labour required for large-scale farming and infrastructure was supplied through collective obligations rather than wage labour or commercial hiring.

Economic exchange in the Inca Empire functioned through Redistribution and barter, lacking markets in the conventional sense. Ethnohistorical evidence suggests that while local barter markets and fairs existed in some regions, the core imperial economy was largely non-commercial. Although later researchers have noted that this portrayal might be somewhat idealised, price-setting market institutions were not a driving force in Inca society. Instead, the state and local authorities (*curacas*) collected agricultural surplus – either as tribute or through communal work – and stored it in vast state warehouses (*qollqas*) for redistribution in times of need or for political ceremonies. Rostworowski (2005) emphasises that reciprocity and the plea-and-gift dynamic underpinned these exchanges. This reciprocal exchange system, described by Rostworowski as the gear of production and distribution in an economy without money, allowed the empire to function smoothly without currency. Goods moved through barter and tribute along hierarchical lines – for example, farmers might send maise to state storehouses, which the state later redistributed as victuals for armies or as famine relief, often accompanied by ritual reciprocity.

### 2.2. On Social and Political Organisation

Inca society was highly stratified and organised into units that facilitated this economic model. At the local level, the *ayllu* (extended kin group) was the fundamental unit of production and social identity. The land was not privately owned but allocated by the community: a couple received a plot (measured in *tupus*) to cultivate at marriage, with allotments adjusted based on family size (additional land for each child). If a family line died out, their land reverted to the *ayllu* for Redistribution, ensuring land circulated to where labour was available. Each *ayllu* ideally possessed lands in various eco-zones (reflecting the vertical archipelago model) and was led by local *curacas* who coordinated communal work and mediated with higher authorities.

Above the *ayllu*, the empire was administratively divided into hierarchical layers: provinces overseen by governors and the four major quarters (*suyus*) under the central rule of the *Sapa Inca* in Cuzco (D'Altroy, 2014). This chain of command enabled top-down mobilisation of resources. A key institution was the *mita*, a rotational labour draft that required communities to supply workers for state projects periodically. Through the *mita*, thousands of commoners might be summoned to build a road, terrace a mountainside, or serve in the army for a season, after which they returned home. Hu and Quave (2020) note that *mita* obligations sometimes blurred into forms of unfree labour – for example, relocated colonists (*mitmaqkuna*) or retained servants (*yanakuna*) who were permanently assigned to state service. Nonetheless, the system was ideologically framed in terms of mutual obligation rather than slavery: the state provided maintenance (food, clothing, security) for those doing *mita*, claiming their service as a form of tribute rather than coerced exploitation.

Andean traditions of *ayni* and *minka* further reinforced social cooperation. *Ayni* refers to direct reciprocity among kin or neighbours, such as helping each other in planting or housebuilding. *Minka* is collective work for communal benefits, like cleaning irrigation canals or communal fields. These practices predated the empire but were co-opted and expanded by the Incas. The *mita* can be seen as a state-wide extension of *minka*, essentially communal labour at the imperial scale. In all cases, the Andean ethos required that labour given was compensated – not in cash, but through the distribution of goods, public feasts, or future reciprocated labour. As Murra and others argue, this created a "give-and-take system" that bound communities together and to the state. Authority was thus exercised through generosity and obligation. In effect, political power in the Inca Empire was inseparable from the management of food and work, a point echoed by Ramírez (2009), who describes Andean leaders' ability "to feed and be fed" as the cosmological basis of their authority. The Incas institutionalised this by building thousands of granaries and storage facilities across the empire, from the highlands to the lowlands. Spanish and *mestizo* chroniclers such as Cobo, Cieza de Leon, Guaman Poma de Ayala, and Inca Garcilaso de la Vega (often cited in the works of Arellano, Murra, Ramírez and Rostworowski) even depicted the Inca storehouses (*qollqa*) graphically in his chronicles, underscoring their ubiquity and importance in governance. The scale of state storage astonished the Spaniards: for example, at the provincial centre of Huánuco Pampa, archaeologists have documented over 4000 *qollqa* buildings, used to stockpile corn, potatoes (dried as *chuño*), quinoa, and other staples for Redistribution.

## 2.3. On Knowledge Transmission and Administration

Operating a vast empire without writing required alternative knowledge transmission and record-keeping means. Inca society relied on an oral tradition for historical and cultural knowledge and physical devices like the *quipu* for administrative data. A *quipu* (Quechua for "knot") was a collection of coloured knotted strings used to encode numeric information – such as census figures, tribute quotas, and storehouse inventories – in a decimal place-value system. Recent scholarship suggests *quipus* could also encode non-numeric details in a more limited way by using distinct sequences as mnemonic prompts for oral narratives. Arellano Hoffmann (2013) provides an example of *quipus* used in early colonial times to record local tribute obligations, indicating that indigenous officials adapted *quipu* record-keeping to the new Spanish administration. This information underscores that *quipu* literacy (handled by specialist *quipucamayocs*) was a tightly guarded skill passed down through apprenticeship. Training was done by practice and memorisation rather than written manuals – an illustration of heuristic learning.

The Inca Empire's knowledge of agriculture, engineering, and medicine was empirical and accumulated through generations. Farmers and artisans operated with a deep understanding of local conditions and learned through trial and error. For example, freeze-drying potatoes into *chuño* – by exposing potatoes to the freezing night air and morning sun and then trampling out the moisture – was a scientific innovation achieved without formal science but through iterative experimentation in high-altitude communities. Similarly, the construction of suspension bridges from woven *ichu* (grass) or the precise cutting and fitting of stone masonry were arts taught by demonstration and sustained by continuous practice. This practical, heuristic mode of knowledge transmission meant that know-how was embedded in social processes (the collective memory of the *ayllu*, or the specialised guilds of workers) rather than in written archives. The *yachachiq* (informal teachers) and elders in each community played a key role in mentoring the young in agricultural cycles, weaving patterns, and ritual practices – effectively ensuring that each new generation inherited the lessons of the past.

Colonial-period accounts hint at how adaptable Inca knowledge systems were. Even after the Spanish conquest, indigenous record-keepers known as *quilcaycamayoc* continued to compile information, sometimes learning to use the Spanish alphabet on paper while drawing on their quipu-honed skills (Burns, 2011). Such continuity shows that Inca administrators had developed a robust oral-visual means of governance. D'Altroy (2018) notes that Inca administrative sophistication – with its census counts, storehouse audits, and hierarchical supervision – rivalled that of contemporaneous literate states. The Incas prove an important point in social science; complex state operations can be managed through non-written, non-monetary mediums, given the proper social organisation and cultural investments in training and memory. Our literature review thus points to a nexus of food, labour, and Knowledge in Inca society, all organised through a principle of cooperative reciprocity under stratified authority.

## 3. METHODOLOGY

This research adopts a historical-analytical methodology grounded in ethnohistorical sources, archaeology, and comparative anthropological theory. The approach is interdisciplinary, combining:

### 3.1. Documentary Analysis

Thanks to Spanish and *mestizo* chroniclers, modern historical interpretations (Rostworowski, 2005; D'Altroy, 2018) have been used to reconstruct the Inca economy and governance systems.

These sources provide qualitative descriptions of Inca practices observed or remembered in the early colonial period, which we critically analyse, considering potential biases. We also incorporate data from archaeological studies (such as those on agricultural terraces, water management, and settlement patterns) to corroborate and elaborate the written accounts.

### 3.2. Comparative Framework

Using economic anthropology and sociology concepts, we interpret Inca institutions through theoretical lenses. Munck's work on Karl Polanyi's notions of Redistribution (2015) and Murra's vertical archipelago model provide the basics. We introduce the new conceptual model – *barter economy with hierarchical cooperation* – as an analytical tool and compare it against existing frameworks, such as the "socialist state" hypothesis of Baudin (1961) or the "administered trade" model of Stanish and Coben (2013). This comparative element helps highlight what was unique about the Inca case versus other pre-modern economies.

### 3.3. Heuristic-Practical Perspective

Methodologically, we pay special attention to evidence of heuristic learning (trial-and-error, pragmatic adaptation) in the Inca record. This approach involves analysing technological and agricultural developments for indications of incremental innovation. For example, we consider agronomic practices (crop rotations, soil amendments, storage techniques) and infrastructural projects as outcomes of collective learning processes rather than top-down scientific planning. By reconstructing how Knowledge might have been generated and transmitted, we align our method with the subject of inquiry – effectively using a *practice-oriented* analytical lens on a practice-based society.

Given the qualitative nature of the evidence, our analysis is interpretative and exploratory. We do not employ statistical analyses, but we do triangulate multiple sources to ensure the reliability of the historical facts presented. The combination of ethnohistorical narrative and modern analysis allows us to infer the underlying "social logic" (in Murra's terms) of the Inca system. All interpretations and theoretical contributions are grounded in documented evidence with citations to primary or secondary sources.

### 3.4. Hypotheses

Based on the literature review and our theoretical framing, the study is guided by the following hypotheses:

1. The Inca Economic System Operated as a Barter-Based Hierarchical Network: We hypothesise that the Inca Empire's economy can best be understood as a non-monetary barter system with centralised, hierarchical coordination. In this model, the exchange of goods and labour was regulated by social obligation and state redistribution rather than market pricing, and cooperation was enforced through a top-down hierarchy (from the *Sapa Inca* down to households) rather than through voluntary market transactions.

2. Heuristic Knowledge underpinned Inca Innovations: We propose that the Incas' achievements in agriculture and engineering were primarily the result of a heuristic, practice-driven accumulation of Knowledge. That is, systematic trial-and-error and adaptation to local conditions, rather than formal written Knowledge or external instruction, produced effective solutions (e.g. terracing, *quipu* accounting, freeze-drying) disseminated through experiential learning in the population.

3. Reciprocity and Redistribution Ensured Social Cohesion: A key hypothesis is that the principles of reciprocity (exchange of gifts and labour) and redistribution (central collection and disbursal of resources) created a stable social contract between the Inca state and its subjects. We may say that these principles were both economic and ideological, providing social cohesion and legitimacy to the political hierarchy. This notion entails those instances of resource allocation (such as famine relief from state storehouses or communal labour for public works) that were perceived as mutual fulfilment of duties, reinforcing loyalty to the state.

These hypotheses serve as lenses through which we interpret the historical evidence. In the following sections, we present the results of our investigation, examining to what extent the evidence supports these propositions and then discuss their implications in a broader context.

## 4. RESULTS AND DISCUSSION

### 4.1. On Food Production and Security in a Diverse Environment

Our examination of Inca food production confirms an elaborate system geared toward maximising output and buffering against scarcity. The empire encompassed at least four major ecological zones (coast, highlands, intermediate valleys, Amazonian fringe), yielding different staples. Rather than trade outputs between independent groups, the Incas politically integrated these zones and moved goods via llama caravans along the state road system. Crops like maise (corn), quinoa, potatoes, and coca leaf were grown in optimal zones and then redistributed. For example, highland communities sent quinoa and *chuño* to the lowlands, while coastal communities sent dried fish or cotton textiles to the highlands – all as part of tribute quotas. The imperial administration closely monitored production: annual censuses of farmers and inventories of herds were taken (facilitated by *quipus*) to assess the resources available. We find strong evidence supporting Hypothesis 1 in this domain; the movement of produce was accomplished through barter-like exchanges mediated by the state. There was no buying or selling of corn in a marketplace; instead, the state would allocate corn from its storehouses to communities with a bad harvest in exchange for their future labour or other products. Such transactions fundamentally differed from market trade – they were embedded in state-administered reciprocity.

The physical infrastructure for food security was impressive. As hypothesised, reciprocity and Redistribution were key – the people's tribute filled local storehouses, and in return, the stored goods sustained the people in times of need. Our findings show that these were strategically placed in various climates to preserve different goods (freeze-dried foods in cold highlands, maise in dry coastal sites, etc.), demonstrating systemic planning for risk mitigation. When droughts or El Niño events caused regional crop failures, the Inca bureaucracy was able to respond by moving stored provisions to the affected area, averting famine – a fact recorded by chroniclers and evidenced by the absence of severe famine in Inca times despite frequent climate oscillations (Antúnez de Mayolo, 2011). This result underscores Hypothesis 3; social cohesion was reinforced by the state's role as guarantor of subsistence.

Additionally, agricultural practices uncovered in the study highlight the heuristic nature of Inca knowledge, which aligns with Hypothesis 2. The Incas employed terrace farming extensively; each terrace was a learning laboratory where farmers adjusted soil depth, water flow, and crop rotation to sustain yields on steep slopes. This method required intimate knowledge of microclimates, which was gained through centuries of local experimentation. One result supporting this is the design of terraced fields in Moray (near Cuzco), believed to be an Inca agricultural research station: concentric terraces there create distinct temperature zones, possibly used to observe crop performance. Similarly, natural freeze-drying and sun-drying of meat (*charqui*) are technologies likely discovered empirically and propagated because they worked reliably in Andean conditions. Nowhere is there evidence of a formal scientific theory guiding these innovations; instead, the Knowledge was transmitted via practice – fathers to sons, mothers to daughters – confirming the hypothesis that practical experience was the teacher in the Inca Empire.

### 4.2. On Social Organisation and Labour Cooperation

Our findings illustrate a finely tuned system of labour allocation orchestrated through social units and state oversight. The *ayllu* fulfilled the role of a communal management unit; within each *ayllu*, families cooperated in farming each other's fields in succession (*ayni*), ensuring everyone's crops were planted and harvested on time. This cooperative labour at the grassroots level meshed with state needs through the *mita* system. We found ample documentation that the *mita* was a taxation system paid in labour. For example, records indicate that each *ayllu* had to send a certain number of men each year to serve in state projects (mining, army, or construction) – a quota determined by provincial governors and tracked by quipu. These men would be rotated so that no community's workforce would be depleted for too long. During their service, the state provided for them from state stocks, confirming that these obligations were conceived in terms of exchange (service given for support received) rather than one-sided extraction. This statement strongly supports Hypothesis 1 in the labour domain; labour was the "currency" of the empire, and its exchange was regulated by a hierarchy of authority but justified through a cooperative ideology. Even large infrastructure projects like the famous royal road system or the fortress of Sacsayhuaman were built not by slaves but by temporary labour conscripts from across the realm, coordinated by engineers and supervisors in a remarkable feat of human organisation (D'Altroy, 2018). The hierarchical cooperation is evident – at the top, the Inca and his administrators planned and requested; at the bottom, commoners executed the work in a spirit of duty, and in between, local chiefs mediated and organised their contingents. The study also reveals that the ethics of reciprocity mitigated social stratification. While the *Sapa Inca* and nobility extracted surplus from the populace, they were expected to redistribute largesse. One outcome observed is the role of state-sponsored feasts (redistributive feasting). After major work projects or during religious ceremonies, the Inca or his governors would host enormous feasts, slaughtering hundreds of llamas for meat and chicha beer to distribute to the workers. These feasts both served to reward labour and dramatise the rulers' benevolence. Belloc and Bowles (2017) suggest that in autarkic regimes (with little external trade), such internal redistributive events help reinforce institutional stability by culturally embedding the norms of cooperation. Our findings align with that view: the Incas maintained stability in part by making the population dependent on and appreciative of the state's provisioning. The cosmological element noted by Ramírez (op. cit..) also appears – the Inca was seen as a paternal figure who fed his people (literally and ritually), which in Andean cosmology legitimised his right to rule.

On the local level, we also note some stress inherent in the system: the *curacas* (local nobility) sometimes struggled between fulfilling Inca demands and maintaining the welfare of their kin groups. Rostworowski (2005) questioned whether the "gifts" given by the Inca were truly sufficient to offset the loss of tribute that *curacas* had to surrender to the state. There is evidence that some local lords were aggrieved by having their best lands and workers requisitioned. However, the Incas often co-opted these elites through inclusion in the imperial administration and allowing them privileges (such as access to exotic goods or keeping a portion of tribute). Overall, the results indicate that the mechanism of hierarchical cooperation, while not devoid of coercion, was largely effective in mobilising the populace with minimal open rebellion – a testimony to how deeply ingrained reciprocity was in the social fabric.

### 4.3. On Knowledge Systems and Communication

Investigation into administrative records and knowledge transmission validates the role of quipu-based accounting and oral communication as the twin pillars of Inca information management. As described in the literature, the *quipu* system was ubiquitous in recording vital statistics: warehouse tallies, census data, labour service records, and possibly calendrical or ritual schedules. We found specific instances (from early colonial reports) of *quipucamayocs* explaining their *quipus* to Spanish officials, indicating that *quipu* cords accurately encoded data like population numbers and stored quantities (Arellano Hoffmann, 2013). Hypothesis 2 underscores that the Incas developed their own pragmatic knowledge device – an indigenous "database" – through heuristic means. The decimal structure of *quipu* (units, tens, hundreds, etc., indicated by knot positioning) shows a clear understanding of place value, which might have evolved from using counters on a board (the *yupana* abacus) in conjunction with knotted strings. Such an invention was likely the result of practical problem-solving; faced with administering surpluses and obligations across an empire, the need to remember and compute large numbers led to this innovative record-keeping system. It is a strong example of how necessity and experience drove intellectual development in the Andes.

The results also highlight the role of narrative and oral tradition. Preserving history, technical knowledge, and culture without writing relied on oral specialists. The Inca had official "memorisers" – the *amautas* (philosopher-teachers) and *haravicus* (bards) – who were tasked with learning lore and teaching it to younger generations. This, too, was an organised form of knowledge transmission, albeit non-written. We found that rituals and annual ceremonies often functioned as educational forums: for instance, during planting festivals, elders would recite the origin of specific crops or past episodes of famine and relief, thereby instilling practical tips and moral lessons in the community. The heuristic-practical perspective is evident here; knowledge was inseparable from practice and usually taught in context.

Finally, our study discussed how the Incas' lack of writing influenced their governance style. We observed that it necessitated face-to-face communication at every level of administration. Authorities had to travel or send messengers (*chasquis*) to convey orders and check on distant communities. This ironically strengthened control, as it forced a continuous presence of the state in local affairs (through visits, inspections, and ceremonies) rather than remote rule by decree. The reliance on memory could also limit the complexity of regulations – administrative directives had to be simple enough to be memorised and passed on. Indeed, the legal and economic norms in the empire were often codified in ritual or oral formulae (for example, a set phrase recited when distributing lands or calling up *mita* workers), which made them easier to remember. This finding reinforces Hypothesis 3 insofar as the interactive nature of communication (personalised, reciprocal) likely enhanced social cohesion; people dealt with known messengers and familiar officials rather than impersonal edicts. In summary, the results across these domains support the notion that a *hierarchically coordinated yet cooperative* system sustained the Inca Empire. Material evidence and historical accounts consistently point to an economy of barter and reciprocity managed through a powerful state apparatus. Knowledge was disseminated and preserved through doing, seeing, and remembering rather than abstract recording. The Incas achieved a remarkable alignment of practical know-how with social organisation, vindicating the idea that complex societies need not fit the model of market economies or literate bureaucracies to thrive. Figures 1 and 2 help to illustrate our ideas.

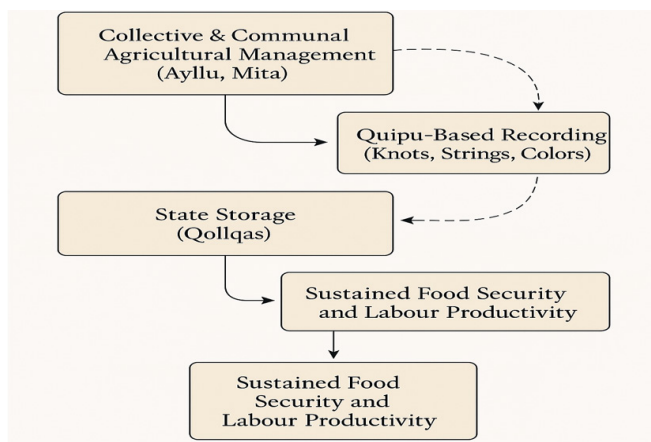

**Figure 1.** Flow diagram of Inca agricultural and economic organization

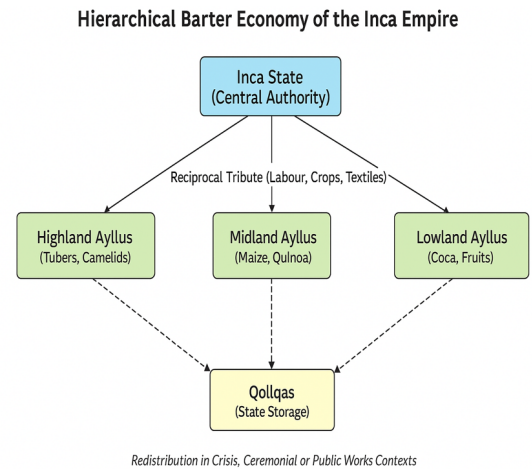

**Figure 2.** Hierarchical barter economy of the inca empire

### 4.4. Discussion

The Inca Empire presents a compelling model of how social complexity can emerge outside market institutions or written bureaucracy. Our findings support the concept of a barter economy structured by hierarchical cooperation, offering broader insights into economic anthropology and comparative institutional theory.

Rather than relying on monetary incentives or price mechanisms, the Incas coordinated food production, labour, and distribution through centralised planning and embedded reciprocity. This aligns with Polanyi's redistribution framework yet adds nuance: Inca exchange combined vertical authority with local-level obligations, forming a stratified system of mutual provision. These dynamics challenge the assumption that money is a prerequisite for managing large-scale economies.

Innovation in the Inca world followed a practice-based logic. Technological solutions like terracing, freeze-drying, and *quipu* record-keeping emerged from cumulative experimentation and collective memory, not formal science. Oral transmission, embodied teaching, and role specialisation sustained knowledge across generations. While these systems proved effective for empirical problem-solving, they also imposed cognitive limits on abstraction and theoretical development.

Comparative examples underscore the distinctiveness of the Inca model. Like ancient Egypt or Angkor, the Incas mobilised labour and redistributed staples without money. Yet their near-total reliance on non-market mechanisms, shaped by geographic isolation and ecological diversity, set them apart. They constructed internal resilience through local self-sufficiency and state provisioning, reinforcing institutional stability via moral obligation rather than commercial transaction.

The interplay between economy and power was central to Inca governance. Control over staple resources allowed the state to secure loyalty and maintain order. Redistributive feasts, material support during hardship, and symbolic generosity affirmed political authority while meeting subsistence needs. This welfare-oriented structure was effective, but vulnerable to disruption, as seen during the Spanish conquest, when demographic collapse and institutional breakdown dismantled the system's delicate balance.

The Inca Empire's organisational model offers rich insights into alternative pathways of social complexity. The evidence strongly supports our concept of a barter economy based on hierarchical cooperation, and here, we discuss its significance relative to broader theoretical and comparative contexts. Notably, modern economic theory has started to appreciate such patterns; for example, Roth (2018) discusses how even market design can involve matchmaking without money in certain circumstances (like kidney exchanges), analogous in some sense to how the Inca state "matched" surpluses with deficits without a price mechanism.

The Incas particularly stood out for achieving empire-wide integration; many other examples (like the Maya or Mesopotamia) had markets alongside redistribution, whereas the Incas leaned almost entirely on non-market mechanisms. This uniqueness may relate to their isolation (no adjacent comparable states to trade with) and ecology, since the Andes made transport difficult, so local self-sufficiency was crucial. Belloc and Bowles (2017) argue that long-term autarchy can entrench localised institutions and cultural norms– the Incas exemplify this, having developed strong internal institutions partly due to minimal external trade. Stanish and Coben (2013) note that marketplaces in the Andes were likely more on the fringes or for luxury goods. Our discussion clarifies that acknowledging minor market activity does not contradict the overall model; rather, it shows the Incas tolerated small-scale barter markets insofar as they did not undermine the state's dominance over staple distribution.

From a social science perspective, the Inca model also provides a case study of the relationship between power and economy. The control over food–as power over life–was a cornerstone of Inca political strategy. The Inca elite could coerce and persuade their subjects by monopolising the storage and redistribution of staples. This resonates with theories in political anthropology about staple finance vs. wealth finance: In Inca society, wealth (precious metals, fancy goods) was less central than staple finance (grain, cloth, etc.) in exercising power. We see that the Incas also extracted luxury items (fine textiles, gold for state use), which were primarily used in ceremonies or as rewards for loyalty, not traded commercially. The key theme is that material well-being and social order were connected – the Incas essentially ran a welfare state in exchange for obedience. Such arrangements can foster stability but are vulnerable to shocks; if the state fails to deliver (e.g., due to catastrophe or mishandling), the social contract frays quickly. Indeed, when the Spanish invaded and disrupted the Inca state, local groups did not uniformly resist – some saw an opportunity to escape heavy labour duties. The catastrophic population decline after the conquest (Livi-Bacci, 2006) broke the back of the system by removing both the taxpayers and the caretakers of the system. This underlines that while the Inca model was highly successful on its terms, it was finely balanced and depended on a closed environment with steady demographic contributions.

Finally, our proposed analytical framework contributes to contemporary sustainable and cooperative economics discussions. Modern societies face questions about organising production and distribution in inequitable, non-extractive ways. When stripped of its historical context, the Inca example offers an experiment: Can large populations be fed and infrastructure maintained without cash, markets, and capitalist incentives? The Incas show it was possible through a moral economy approach built on kinship, obligation, and mutual aid enforced by authority. Of course, translating this to today is infeasible in full, but elements like local time banks, community-supported agriculture, or state grain reserves echo Inca-like solutions. The Incas also present a case where economic aims were explicitly social – to ensure everyone had enough and that the state could mobilise labour for grand projects, rather than to maximise profit or growth. In an era concerned with sustainability, such a perspective is valuable. Turner and Klaus (2020) argue that food and power were fundamentally linked in ancient Andean societies; our findings affirm that and suggest the Incas achieved a degree of food security and resource distribution that many modern states struggle to attain.

## 6. CONCLUSION

This study has analysed the Inca Empire's food production systems, social organisation, and knowledge transmission through a heuristic and practical lens, advancing the concept of a *barter economy based on hierarchical cooperation*. Our findings demonstrate that the Incas operated a complex imperial system without money or writing by relying on structured reciprocity, collective labour, and empirically derived Knowledge. The *ayllu*-based communal economy and state-administered *mita* labour service ensured agricultural abundance and infrastructure development, while extensive state storehouses and redistribution mechanisms safeguarded against famine. The Inca ruling class kept its authority by acting as both organisers and benefactors. Knowledge in farming techniques or the accounting *quipus* was generated and shared through practice, illustrating that innovation can thrive outside formal scientific traditions.

As proposed, the concept of hierarchical cooperation in a barter context encapsulates the essence of Inca socio-economic organisation: a non-market system where cooperation was not egalitarian but orchestrated through rank, and exchange was not absent but conducted without currency. This adds to the academic discourse by providing a concrete model against which to compare other pre-capitalist economies. It reminds us that human societies have devised multiple solutions to the problem of coordination and distribution – the Inca solution emphasised mutual obligations and central planning over individual transactions and competition.

In closing, the Inca Empire illustrates a fascinating alternative pathway of human development – one where *bread and information* were managed without coin or script and where social bonds and state authority wove together an economy of collective well-being. It stands as a testament to the potential of hierarchical yet communitarian organisations and invites us to appreciate the diverse capacities of past societies in solving universal human challenges.

## RECOMMENDATIONS

The significance of the Inca case extends beyond Andean history. They achieved notable sustainability and social welfare feats within their domains, inspiring admiration and further inquiry. However, their system's limits and eventual collapse (under the impact of conquest and disease) also serve as a caution about rigidity and isolation. Future research could build on the heuristic-practical framework applied here by examining how non-market, cooperative economies adapt to stresses or interact with market-based systems. Comparative studies between the Inca and other empires could refine our understanding of when and why different economic logics prevail.